**FRONT MATTER**

**Title**

Applying deep learning to teleseismic phase detection and picking: PcP and PKiKP cases

**Authors**

Congcong Yuan[1] and Jie Zhang[1]

*Congcong Yuan: congcy@mail.ustc.edu.cn*

*Jie Zhang: jzhang25@ustc.edu.cn*

**Affiliation**

[1]Geophysical Research Institute, School of Earth and Space Sciences, University of Science and Technology of China, Hefei, Anhui 230026, P. R. China.

**ABSTRACT**

The availability of a tremendous amount of seismic data demands seismological researchers to analyze seismic phases efficiently. Recently, deep learning algorithms exhibit a powerful capability of detecting and picking on P- and S-wave phases. However, it is still a challenge to process teleseismic phases fast and accurately. In this study, we detect and pick the PcP and PKiKP phases from a Hinet dataset with 7386 seismograms by applying a deep-learning-based scheme. The scheme consists of three steps: first, we prepare latent phase data, which is truncated from the whole seismogram with the theoretical arrival time; second, we identify and evaluate the latent phase via a convolutional neural network; third, we pick the first break of good or fair phase with a fully convolutional neural network. The detection result shows that the scheme recognizes 92.15% and 94.13% of PcP and PKiKP phases. The picking result has an absolute mean error of 0.0742 s and 0.0636 s for the PcP and PKiKP phases, respectively. The performance of the picking algorithm is compared with the traditional approach of STA/LTA. The scheme processes all 7386 seismograms approximately in 2 hours, especially only cost about five minutes on the last two steps.

## INTRODUCTION

It is time- and labor-consuming for seismologists to analyze seismogram and extract seismic phases effectively for in-house research purpose, especially in the face of current voluminous seismic data (Zhang et al., 2003; Ross and Ben-Zion, 2014). Although available algorithms enable a seismologist to process a large amount of data and interpret seismograms reasonably well (Havskov and Ottemöller, 2009; Bormann et al., 2014), the interactive analysis and interpretation require the intervention and evaluation by seismologists. An efficient process for seismogram analysis and interpretation is therefore demanding (Romero et al., 2016).

One important task in the routine procedure of seismic data processing is to identify and determine the first break or onset of seismic phases. Accurate detection and picking are crucial for estimating source or model parameters, such as hypocenters, origin time, focal mechanisms, magnitude, and structural imaging, which are of significance for our understanding and monitoring of earthquake triggering as well as subsurface movements (e.g., Kennett and Engdahl, 1991; Song and Richards, 1996; Li and van Der Hilst, 2010; Oueity and Clowes, 2010; Wang and Zhao, 2017; Ross et al., 2018). During past decades, a variety of algorithms have been proposed to detect and pick the P and S phases. Traditional approaches, such as STA/LTA (Allen, 1978), energy-based (e.g., Han et al., 2009), high-order statistics based (e.g., Yung and Ikelle, 1997), autoregressive methods (e.g., Sleeman and van Eck, 1999), the wavelet transform based (e.g., Zhang et al., 2003), cross-correlation-based (e.g., Molyneux and Schmitt, 1999), neural networks (e.g., McCormack et al., 1993), and others (e.g., Boschetti et al., 1996; Mousa et al., 2011), have been reviewed and evaluated by Akram and Eaton (2016) and García et al. (2016).

In recent years, several supervised learning algorithms, including deep neural network (Bengio, 2009), recurrent neural networks (e.g., Hochreiter and Schmidhuber, 1997), convolutional neural networks (e.g., LeCun et al., 1998), and generative adversarial networks (e.g., Goodfellow et al., 2014), have emerged as effective learning approaches that widely employed in a range of fields, such as human vision (Ba and Kavukcuoglu, 2014; Mnih et al., 2015), natural language processing (Bottou, 2014; Xu et al., 2015), and speech recognition (Bahdanau et al., 2015). The survey on the overview of these algorithms can be referred to (Bengio et al., 2015; Schmidhuber, 2015). Several algorithms have been currently modified and extended to detect earthquake (Perol et al., 2018; Li et al., 2018; Mousavi et al., 2018; Zhang and Yuan, 2018; Devries et al., 2018; Ross et al., 2018) and pick the first break or onset of P- or S-wave phase (Chen, 2017; Zheng et al., 2017; Ross et al., 2018; Zhu and Beroza, 2018; Chen et al., 2019). Compared with conventional techniques, deep learning based methods demonstrate promising performance in the applications of seismic detection and picking.

For the methods mentioned above, they are developed exclusively for the detection or picking of P- and S-onsets of seismograms. However, they are seldom introduced particularly for teleseismic phases in seismograms. These teleseismic phases are very important for seismological researchers, since they are often used to investigate the deep subsurface structures. For instances, reflected PmP or SmS phases are useful to estimate Moho depth (Richards-Dinger and Shearer, 1997); reflected PP or SS phases are utilized to define 410 and 660 km discontinuities (Shearer and Masters, 1992); Creager and Jordan (1986) used PKPdf and PKPab phase times to infer heterogeneity in the core and mantle sides of the core-mantle boundary; and PcP and PKiKP phase times are used to study topographic variations of the inner core boundary (Tian and Wen, *2017*). It is difficult to identify and extract these teleseismic phases from the seismogram directly, because their signal-to-noise ratios are relatively low and sometimes distorted by other phase signals or noises. As previous studies (e.g., Kennett and Engdahl, 1991), we use theoretical traveltimes to define a searching time window, in which we identify and pick the target phases with our human eyes. Hence, intensive time and labors are consumed on these tedious processes to retrieve the target phases from tons of seismic data before conducting the subsequent researches.

In this study, we propose a scheme to detect and pick teleseismic phases in an automated manner. It is mainly developed based upon the convolutional neural networks, for which their efficient capabilities have been exhibited in the image processing (LeCun et al., 1998; Ronneberger et al., 2015). Before using the proposed scheme, we assume that the origin time of the selected event has been known. We design a three-step scheme: first, the origin time of the selected event has been assumed to be known before calculating theoretical arrival times of the target phase. We preprocess the input raw seismogram and cut out a latent phase from the processed seismogram based on the time window defined with the theoretical arrival time; second, we discriminate whether the latent phase is good, fair, or poor with a convolutional neural network; third, we discard the poor phase and pick fair or good phase through a separate convolutional neural network. We shall introduce these three implementations and the associated algorithms in the following section. The proposed scheme is then applied to the detection and picking of the PcP and PKiKP phases on a dataset with 7386 seismograms.

**DATA AND METHODS**

**1. Data preparation**

Figure 1 shows the scheme proposed for teleseismic phase detection and picking. It consists of data preparation (Figure 1a), detection network (Figure 1b), and picking network (Figure 1c). In the data preparation, the raw seismic data are input and preprocessed first. Then, we calculate the theoretical arrival time of the target phase with a 1-D velocity model as well as the known origin time. The arrival time is aligned to the preprocessed data. We next define a time window set before and after the arrival time. Finally, we can obtain a latent target phase, which is truncated and achieved by the time window from the processed data. The length of time window is defined based on the processing experience. A shorter window may omit potential target phase, while a longer window may incur more noises and overburden the computation in the following routines.

For example, in this study we aim to detect and pick the PcP and PKiKP phases, compressional waves reflected off the Earth's core-mantle boundary and inner-core boundary respectively (Figure 2). We input one raw event seismogram into the step of data preparation. Then the seismogram is preprocessed with some fundamental processes as follows: (1) remove instrument response and convolute with WWSP response; (2) remove the mean and trend, and apply a symmetric taper to each end of the data; (3) filter the data from 1 Hz to 3 Hz with a Butterworth bandpass filter. Then we calculate the theoretical arrival times (marked by dashed lines in Figure 1a) of our target phases in PREM (Dziewonski and Anderson, 1981) by using the TauP toolkit (Crotwell et al., 1999). We next align them on the corresponding waveforms to truncate the potential phases with a given 20 s' window (marked by a red rectangle in Figure 1a) before and after the theoretical arrival time (marked by the dashed line). These truncated waveforms are viewed as latent PcP and PKiKP phases, which will be further evaluated by a detection network routinely.

## 2. Detection network and training

Figure 1b shows the detection network architecture, which is built upon the convolutional neural network (CNN) that is first proposed by (LeCun et al., 1998). CNN is a powerful deep learning algorithm to effectively extract and recognize the feature of the target object in daily life. Similarly, we cast the target phase selection as a problem of object detection and classification. We build a detection network architecture based on the CNN to detect and classify the latent phases in this study. The input waveform is the latent seismic phase defined as a vector of waveform time samples. Here we fix the number of time samples as 2000. Before applying the network, we have to train the network for detection and classification. During the network training, we perform our body processing by a feed-forward stack of four convolutional layers, to extract useful features of seismic phases, followed by one fully connected layer that outputs the possibilities of three classes, good, fair, and poor. The filter number, size, and max pooling size are denoted in the network architecture. We further optimize the network parameters by minimizing a cross-entropy loss function on a dataset with $N$ training samples,

$$\ell = \frac{1}{N} \sum_{k=1}^{N} p_k \log\left(q_k\right),$$

(1)

$$p = \begin{cases} 2, & good \\ 1, & fair \\ 0, & poor \end{cases}$$

The loss function measures the average discrepancy between the predicted distribution $q$ and the class probability $p$ for all $N$ training samples. We apply the ADMM (Boyd et al., 2010) optimization method with the default learning rate of 0.0001 to minimize the loss function, and update the convolutional kernels and weights during each epoch. To mitigate the overfitting, we also add a dropout, a regularization technique (Srivastava et al., 2014), followed by each max pooling or dense layer. For this study, we select 1000 events recorded by Hinet (Okada, et al., 2004). These seismic events are distributed in the region between 25-50° N in latitude and 110-150° E in longitude. We mark this region out according to Figure 3 in (Tian and Wen, 2017). After the preprocessing, we manually check the quality of each seismogram and retain a total of 10000 seismograms. During the detection network training for PcP or PKiKP phases, we split all seismograms into two parts: a training dataset with 6000 phases (2000 good, 2000 fair, and 2000 bad phases) and a validation dataset with 3000 phases. The validation dataset are utilized to guarantee and evaluate the performance of the training process. Figure 3a exhibits some data samples for the network training. After training the proposed network, we show the validation losses and the corresponding accuracies over 100 epochs in Figure 4a. Blue and green lines represent the training results for PcP and PKiKP phases, respectively. From the accuracy curves, we recognize that the detection accuracies of PcP and PKiKP reach to approximately 92% and 94%. During the training process, we observe that loss curves have some fluctuations for the over-fitting problem to some extent. L1/2 regularizers as well as early stopping approach are suggested to further mitigate the over-fitting issue. The training history indicates that the network is able to extract useful features or patterns from these latent phases and hence it can identify the validation samples effectively.

## 3. Picking network and training

After the detection process, we have checked whether the latent target phase is good, fair, or poor. We continue applying the picking network on the good or fair phase while skipping this step if the phase is poor. We propose a picking algorithm herein to obtain the first break or onset of the phase. Figure 1c shows the proposed network architecture for the phase picking, which is inspired by one fully convolutional network to segment cell structures with high resolution (Ronneberger et al., 2015). Analogous to cell borders, the first break is understood as the border of the seismic phase. We attempt to accurately depict the border of the seismic phase with the modified fully convolutional network. The network consists of a contracting path (left side) and an expansive path (right side). The contracting path follows the typical architecture of a convolutional network. Two convolutions are repeated in each step, followed by a rectified linear unit (ReLU) activation and a max pooling operation for down-sampling. In the expansive path, every step includes an up-sampling of feature map followed by a convolution, a concatenation with cropped feature map, and two convolutions, each followed by a ReLU. The contracting path is able to fast extract features of the first break of the seismic phase and the symmetric expanding path can yield a high-resolution pick map that contains the first break of the seismic phase we desire. The kernel and pooling sizes are $3 \times 1$ and $2 \times 1$ at each step. We also mark the filter number in the architecture.

We show some training samples for the picking network in Figure 3b. The seismic phases and the corresponding pick maps can be observed. We manually pick the first break of the seismic phase and represent it probabilistically with a Gaussian distribution with a standard deviation of 20 milliseconds in a pick map. The use of the Gaussian distribution can help the picking network to facilitate convergence (Ronneberger et al., 2015). The seismic phase and pick map will be regarded as the data and label input into the picking network. We utilize 2000 good and 2000 fair phases as well as their pick maps for the network training and remain other 1000 good and 1000 fair phases for the validation of the picking network training. We optimize the network parameters by minimizing the cross-entropy loss function as equation (1) by applying the ADMM (Boyd et al., 2010) optimization method with the default learning rate of 0.0001, and update the convolutional kernels and weights during each epoch. We train the picking networks for PcP and PKiKP phase, respectively. Figure 4b displays the validation losses of PcP and PKiKP picking networks over 100 epochs. Both loss curves in Figure 4b can fast converge to approximately zero over decades of epochs, which demonstrates that the picking networks are effectively trained and well prepared for the prediction of validation samples.

**RESULTS**

After setting up the processing scheme, in which both the detection and picking networks have been trained effectively, we apply it to a testing dataset with 7386 seismograms, which is input in the scheme one by one. All seismograms are processed automatically. The computation cost is within 1 s on the processing of each seismogram. The last two steps only cost tens of milliseconds in the scheme. To evaluate the prediction results, we manually spend several days detecting and picking PcP and PKiKP phases of these seismograms. These manual detected and picked results are viewed as true results. After comparing predicted and true results, we display the prediction results of PcP and PKiKP detection networks statistically in Figure 5. We accurately predict 2566 PcP and 2825 PKiKP phases with high quality (both fair and good phases) from 7386 seismograms, respectively. To measure the performance of the detection results, we separately enumerate the true negatives (TN), true positives (TP), false negatives (FN), and false positives (FP). The recalls (fraction of good and fair phases detected correctly, $TP/(TP+FN)$) and precisions (fraction of detected phases that are good and fair phases, $TP/(TP+FP)$) are both calculated for three classes of PcP or PKiKP predicted results. Both Figure 5a and b show that the recalls and precisions of good phases are above 95%, while those of fair phases are about 83%, relatively lowest in three classes, especially for their recalls. Compared to good and poor phases, fair phases are not quite clear to be defined manually as training samples. In other words, some ambiguities in definitions may affect the network training to distinguish good phases correctly.

We exhibit one picking result or map for PcP and PKiKP phases, respectively, in Figure 6. The pick map from the picking network shows the probabilistic distribution of the first break. We also calculate the pick map with normalized STA/LTA ratios by using the STA/LTA algorithm (Allen et al., 1978). Some example results, along with predicted ones by the picking network, can be observed in Figure 6. We recognize that both the picking network and STA/LTA algorithm produce accurate picks of first breaks for good PcP and PKiKP phases. However, there are a lot of noises or fluctuations on the normalized STA/LTA ratio maps. When phase data with low quality, such as fair PcP and PKiKP phases, these strong fluctuations may result into large uncertainties on picking results. We can observe that the maximum STA/LTA ratios from STA/LTA algorithm deviate from the manual picks (denoted by red lines) while those from the picking network have a good consistent with the red lines. We extract the maximum values on these maps and view them as the predicted first breaks of phase data. We extract all predicted first breaks and calculate the corresponding picking errors by the comparison with precise manual picks. The picking error is defined as the absolute difference between predicted and true first breaks. We statistically count the phase numbers picked at different levels of errors in time for the picking network and STA/LTA algorithm. We show the predicted results of PcP phases in Figure 7a and PKiKP phases in Figure 7b. For PcP or PKiKP phases, we observe that most errors of the picking network are distributed within 0.1 s. However, those errors of the STA/LTA algorithm are larger than 0.15 s. We also calculate the absolute mean picking errors of first breaks predicted by the picking networks. The errors of PcP and PKiKP phases are 0.0742 s and 0.0636 s, respectively. The sampling rate of all seismograms is 0.01 s. Hence, the picking errors are both within 10 time samples.

The first-break time difference between two seismic phases recorded by different receivers (or seismic interferometry) are often used in some studies (e.g., Shearer and Masters, 1992; Song and Richards, 1996; Tian and Wen, 2017). Under this circumstance, the time difference between two first breaks is usually assumed to be equivalent to that between two first peaks as denoted in Figure 8. The first peak means the first peak or trough followed by the first motion in this study. Since the first breaks of seismic phases are often destroyed and contaminated by noises or other phase signals, first peaks are easier to be identified and picked than first breaks. Time difference would be more precise if we pick the first peaks instead of first breaks. To predict the first peak of seismic phase, we label the first peaks rather than first breaks and train separate picking networks as the way mentioned in the previous section. We find that it is feasible for the picking network to predict the first peak, and it is easier for the picking network to predict the first peak than to predict the first break. It is resulted from the first peak more clearly identified than the first break. We manually check and pick the first peaks of all testing seismic phases. We then statistically count the amount of seismic phases picked at various levels of time errors in Figure 9. We also recognize that the prediction errors of both PcP and PKiKP phases are mainly distributed within 0.05 s. The absolute mean picking errors of PcP and PKiKP phases are 0.0441 s and 0.0392 s, respectively, which are obviously better than that of first breaks.

In order to examine the effects of training samples on the predicted accuracies, we utilize different numbers of the training data to train the detection and picking networks. For each training, the same testing samples are used to evaluate the performance of the detection or picking network. In Figure 10, the upper two lines represent the detection accuracies of the same 3000 latent phases over different numbers of training samples. And the lower two lines denote the mean picking errors of the same 2000 phases over different numbers of training samples. We observe that both PcP and PKiKP detection accuracies improve, and both PcP and PKiKP picking errors decrease as the training samples increase, especially when the amount of training samples becomes from 500 to 1000. It indicates both detection and picking networks are accessible to more useful information from training samples and help to improve their prediction accuracies.

**DISCUSSIONS**

In the proposed scheme, we implement the data preparation to obtain the latent phases before detection and picking of seismic phases, instead of directly detecting phases from the seismograms, since it is hardly possible to straightforward identify the target phases from the seismograms. Besides, it is physically meaningful to pre-detect the latent phases based on the theoretical arrival times of target phases calculated in a 1-D velocity model, even though sometimes it may neglect some potential target phases.

The computational cost of the proposed scheme mainly depends on training cost on the detection and picking networks. In this study, we spend approximately seven hours on the training of detection and picking networks using a single standard CPU processor. It would be faster if using a GPU processor. Upon the detection and picking networks are prepared well, we can apply the scheme to detection and picking target seismic phases from raw seismograms. The scheme can process all testing seismograms in about two hours, especially only spend approximately five minutes on the detection and picking steps, which indicates the scheme enables seismic researchers to process the seismic phases in a surprising speed compared to the manual processing.

The picking network can directly process the latent phases achieved from the step of data preparation without the intermediate step of the phase detection, and the removing of the detection step may simplify the scheme and enhance the processing speed. However, it may be not wise because skipping the picking step may make the picking network produce results on all latent phases, including signals and noises. Hence it may cost much time and labors to qualify and select the first break predictions on signals. As the introduced scheme, the detection step can filter the bad phases or noises and select out the right phases with high quality. We have demonstrated that three steps consisted of this scheme are efficient to detect and pick the PcP and PKiKP phases. In the detection network, the output result is identified as a good, fair, or poor phase. Some example samples are shown in Figure 3a. We can recognize that the feature of good phases are prone to be identified and captured because their signals are relatively strong, however, sometimes it is not clear to discriminate the poor or fair phases, thereby which may cause ambiguity and overfitting in the detection network. It is also the reason why the recall and precision of both predicted poor and fair phases are lower compared to the third class in Figure 9, especially for the fair phases. To mitigate the overfitting during the training, we add the dropout layer in the detection network. We further suggest that the regularization term or early stopping measures can be used in the detection network to reduce the overfitting. It is noted that the detection network identifies and classifies the latent phases only in terms of their waveform quality. As for some special cases, for example, the phase is overlapped by other phase signals, the network cannot work in this situation. To handle these situations, it may demand the examination from the researchers or experts.

It is flexible for the picking network to predict the first break or peak of seismic phase, which depends on the definition of the training labels. Because the signal of the first peak of seismic phase is stronger than that of the first break, we find that the picking network can produce more precise picking results on the first peak than the first break. For some depth phases, such as PcP and PKiKP, it is difficult to identify and pick their first breaks, because their arrivals are usually immerged in other signals or noises, it may result in some uncertainties in picking results. Considering that the time difference of first breaks are often utilized in the associated researches, and the difference of first breaks is assumed to be equivalent to that of first peaks, we can train the picking network to predict the first peaks rather than first breaks, although the picking network is able to predict the first breaks. We have shown and analyzed the first break predictions in the previous section. We also recognize that good seismic phases have higher signal-to-noise ratio considerably than fair seismic phases. These good phases can produce more precise first-break picks than fair phases. In our cases, the absolute mean error is about 0.02 s for good PcP or PKiKP phases. The predicted picking results would become better with more available training samples, which may be inferred from Figure 10. Additionally, manual picks with high precision may contribute to the high efficiency of picking network.

**CONCLUSION**

We propose a scheme that enables high-efficiency detection and picking of teleseismic phases of interest. The scheme consists of data preparation, detection, and picking networks. The step of data preparation preprocesses the input raw seismogram and generates the latent seismic phase truncated from the seismogram with the theoretical arrival time. Both detection and picking networks in the following steps are proposed based on convolutional neural networks, which need to be trained before processing the latent phases. With several thousands of training samples for detection and picking networks, the scheme can be activated and applied to detecting and picking the latent phases. The application to PcP and PKiKP phases demonstrates that the scheme can fast and effectively detect and pick the potential fair and good phases, which can significantly reduce the processing period of teleseismic phases. The scheme is not limited to specific phases and can be extended to detect and pick other teleseismic phases in practice. The trained networks can also be applicable to seismograms recorded in other regions, meanwhile, they can be continually trained and fine-tuned with seismograms from other areas.

**DATA AND RESOURCES**

The data that support the finding of this study are available from the National Research Institute for Earth Science and Disaster Resilience (NIED), Japan, under its data policy.

**ACKNOWLEDGEMENTS**

**LIST OF FIGURE CAPTIONS**

**Figure 1.** The scheme proposed for teleseismic phase detection and picking. The scheme consists of three steps: (a) preparation of latent phase data; (b) detection and classification of three levels of seismic phase data; (c) identification and picking of first break or peak of seismic phase data.

**Figure 2.** Raypaths of PcP and PKiKP phases. PcP (blue) and PKiKP (green) are compressional waves reflected off the core-mantle boundary (CMB) and inner core boundary (ICB), respectively. The red star and black triangle represent a seismic source and receiver at one example epicentral distances of 30°.

**Figure 3.** Example samples for the training of networks, including (a) phase data and labels for the detection network and (b) phase data and labels for the picking network.

**Figure 4.** (a) Validation loss and accuracy curves over 100 epochs during the detection network training; (b) Validation loss curves over 100 epochs during the picking network training.

**Figure 5.** Prediction, recall, and precision of three classes for the 1000 (a) PcP- and (b) PKiKP-phase testing results via the trained detection networks.

**Figure 6.** Four examples of first-break picking results of the tested PcP and PKiKP phases using the picking networks and STA/LTA algorithm. The red line denotes the first break of seismic phase that is picked manually.

**Figure 7.** First-break prediction results of PcP and PKiKP phases. (a) Comparison between the PcP-phase picking results from the picking network and STA/LTA algorithm. (b) Comparison between the PKiKP-phase picking results from the picking network and STA/LTA algorithm.

**Figure 8.** Two examples of first-peak picking results of the tested PcP and PKiKP phases using the picking networks. The red line denotes the first peak of seismic phase that is picked manually.

**Figure 9.** First-peak prediction results for PcP and PKiKP phases via the picking networks.

**Figure 10.** The effects of training samples on the detection accuracy and mean picking error of 100 testing samples.

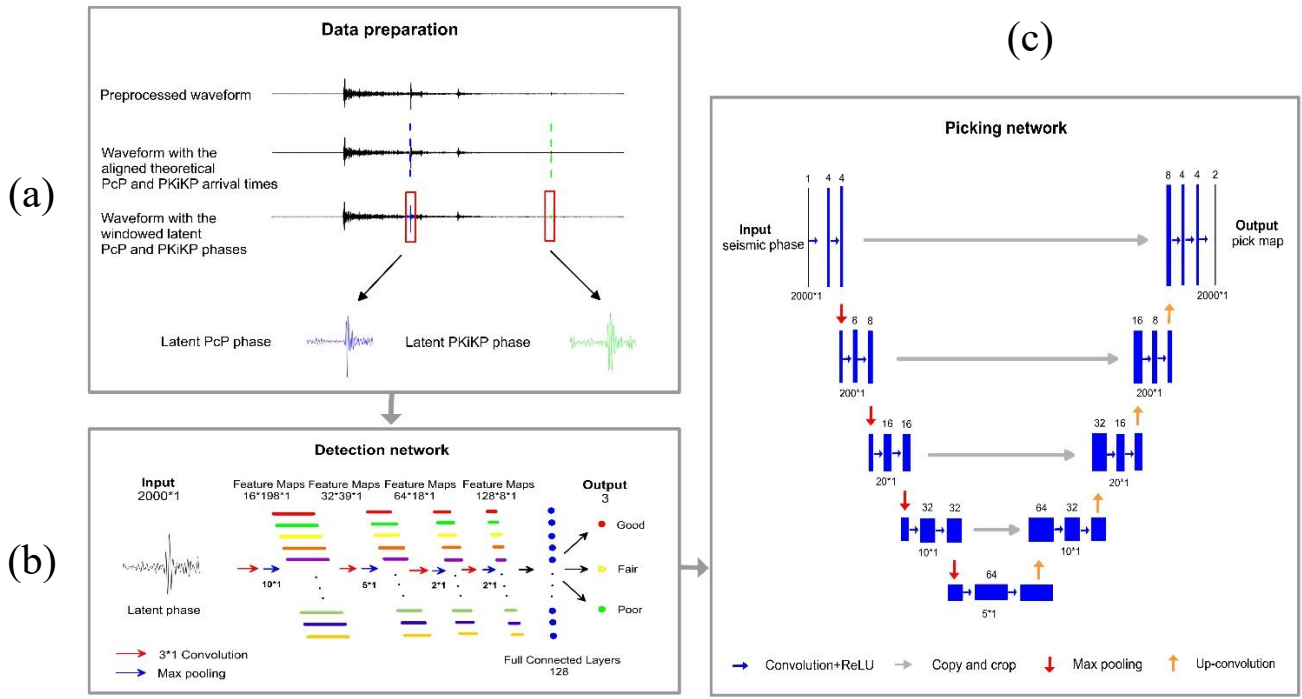

**Figure 1.** The scheme proposed for teleseismic phase detection and picking. The scheme consists of three steps: (a) preparation of latent phase data; (b) detection and classification of three levels of seismic phase data; (c) identification and picking of first break or peak of seismic phase data.

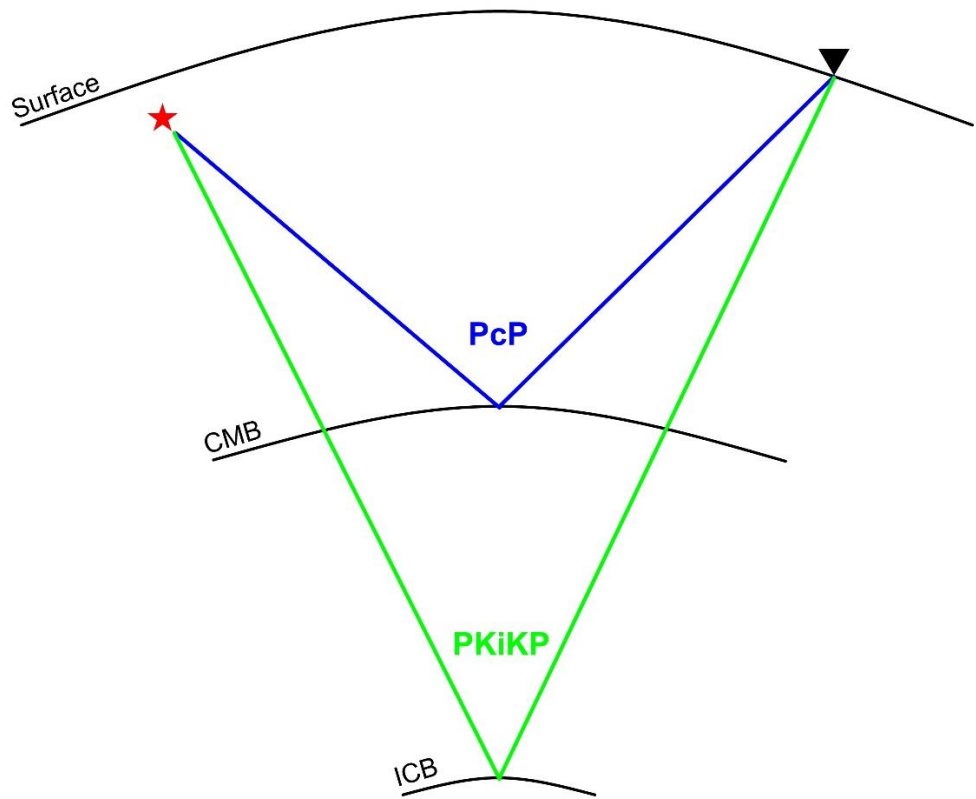

**Figure 2.** Raypaths of PcP and PKiKP phases. PcP (blue) and PKiKP (green) are compressional waves reflected off the core-mantle boundary (CMB) and inner core boundary (ICB), respectively. The red star and black triangle represent a seismic source and receiver at one example epicentral distances of 30°.

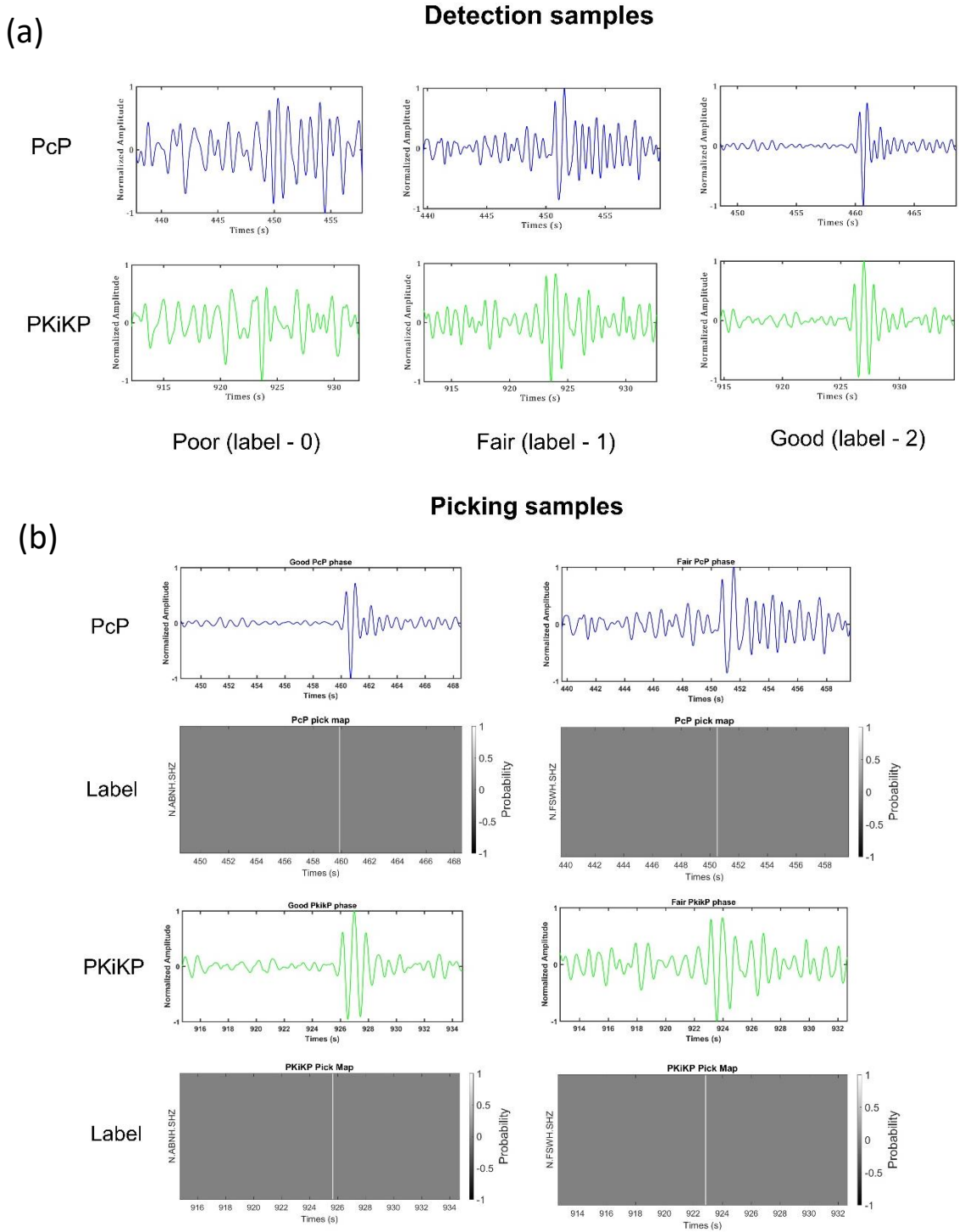

**Figure 3.** Example samples for the training of networks, including (a) phase data and labels for the detection network and (b) phase data and labels for the picking network.

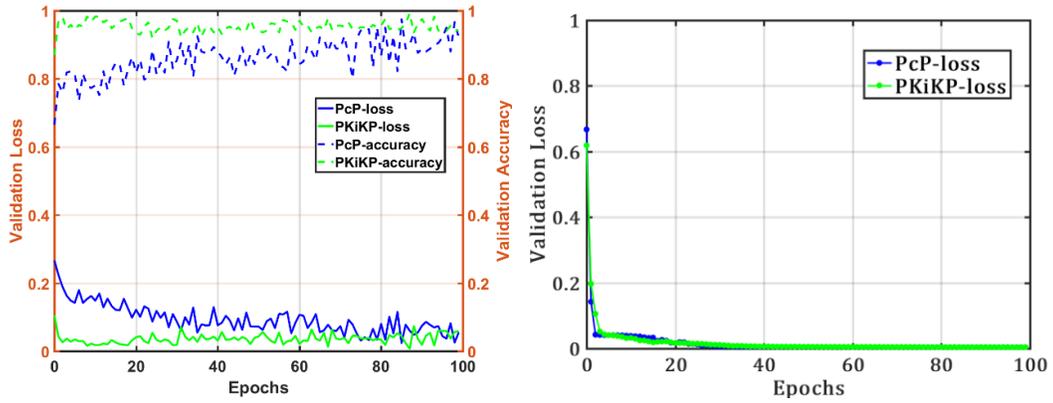

**Figure 4.** (a) Validation loss and accuracy curves over 100 epochs during the detection network training; (b) Validation loss curves over 100 epochs during the picking network training.

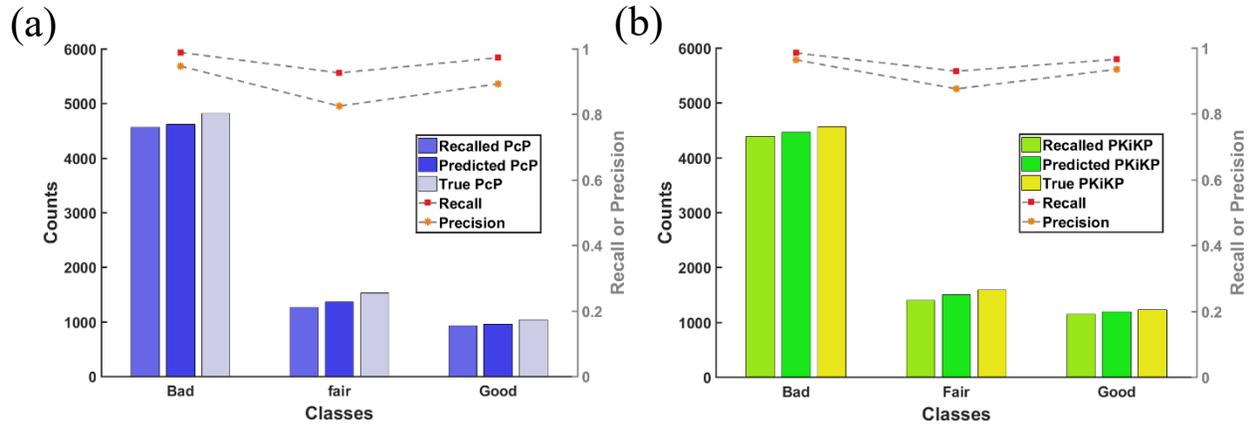

**Figure 5.** Prediction, recall, and precision of three classes for the 1000 (a) PcP- and (b) PKiKP-phase testing results via the trained detection networks.

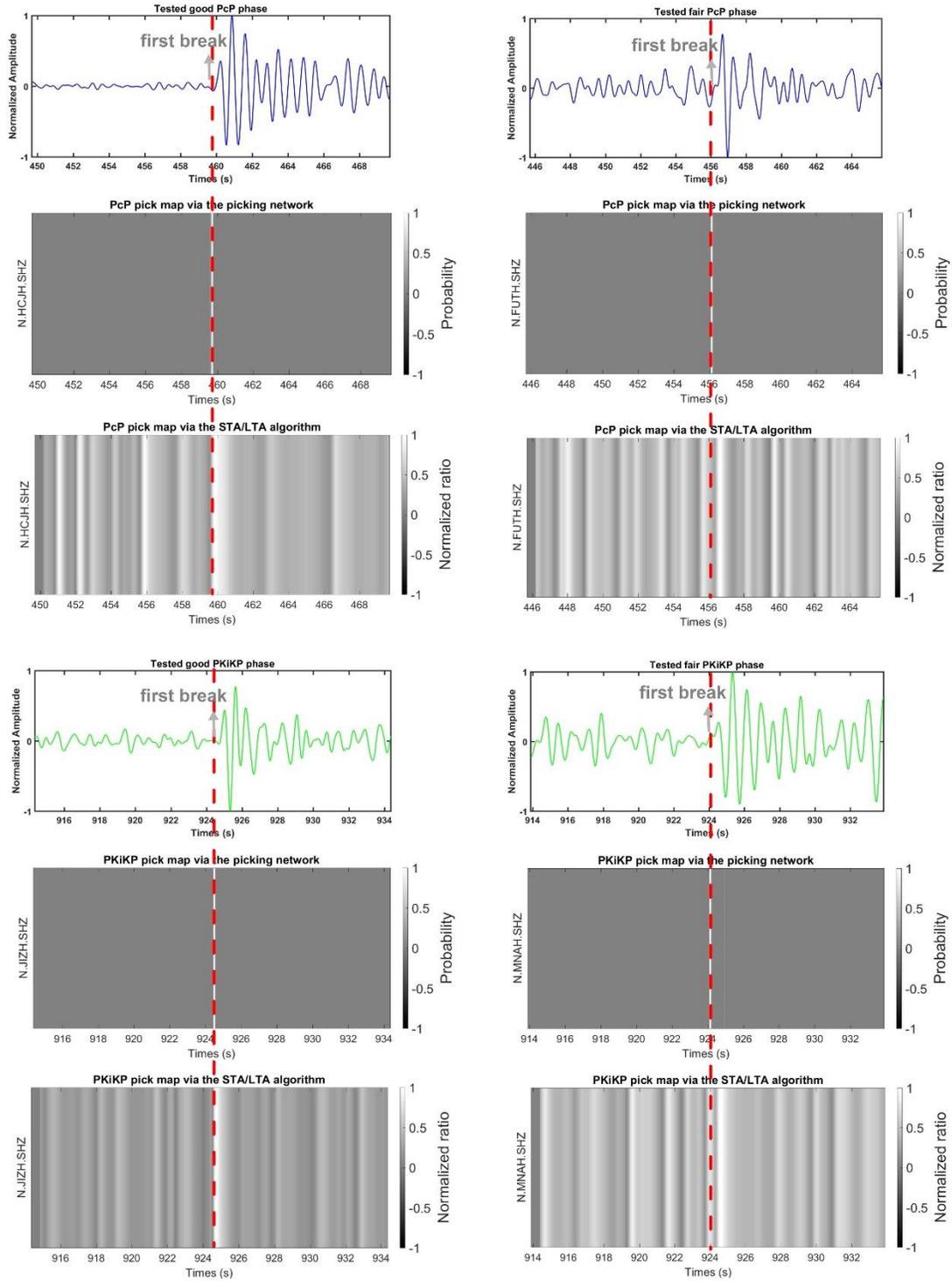

**Figure 6.** Four examples of first-break picking results of the tested PcP and PKiKP phases using the picking networks and STA/LTA algorithm. The red line denotes the first break of seismic phase that is picked manually.

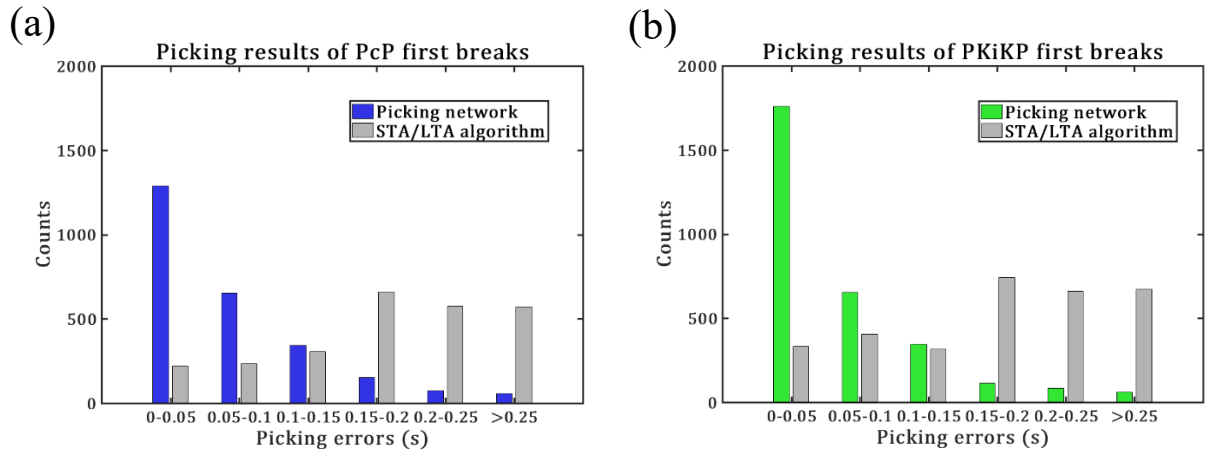

**Figure 7.** First-break prediction results of PcP and PKiKP phases. (a) Comparison between the PcP-phase picking results from the picking network and STA/LTA algorithm. (b) Comparison between the PKiKP-phase picking results from the picking network and STA/LTA algorithm.

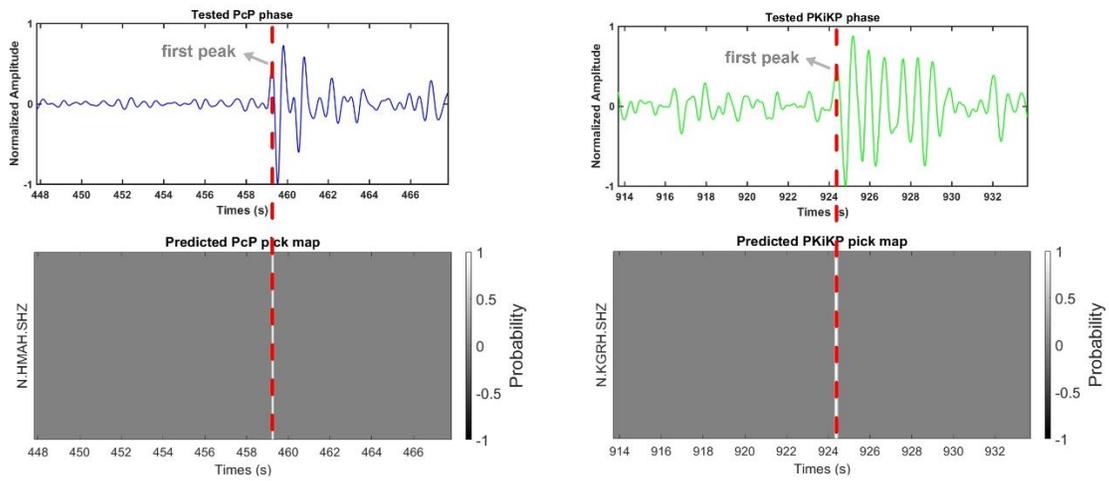

**Figure 8.** Two examples of first-peak picking results of the tested PcP and PKiKP phases using the picking networks. The red line denotes the first peak of seismic phase that is picked manually.

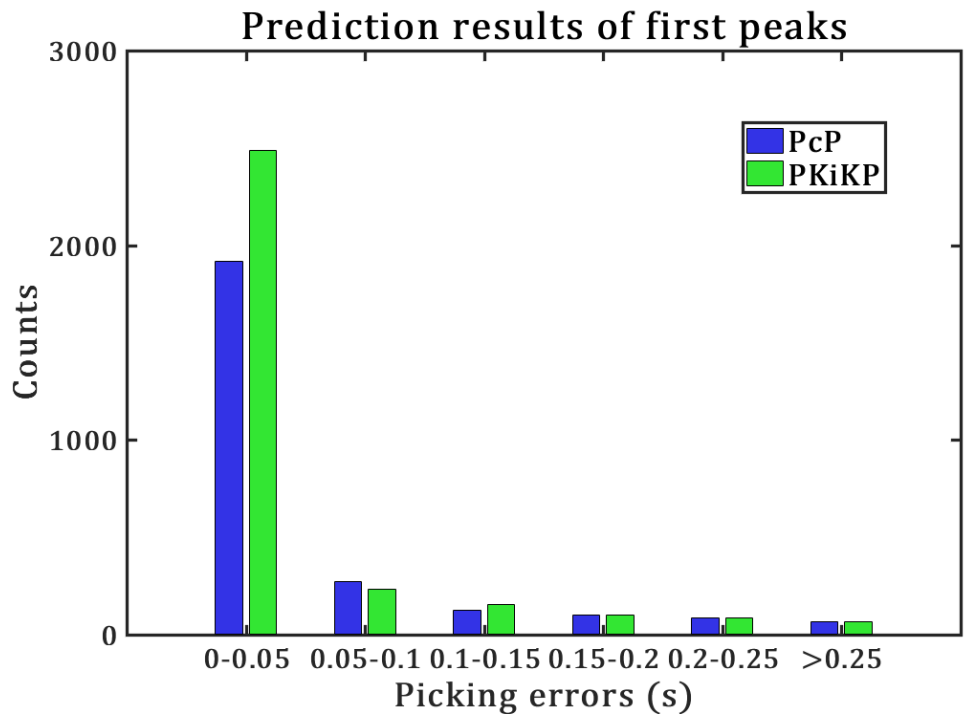

**Figure 9.** First-peak prediction results for PcP and PKiKP phases via the picking networks.

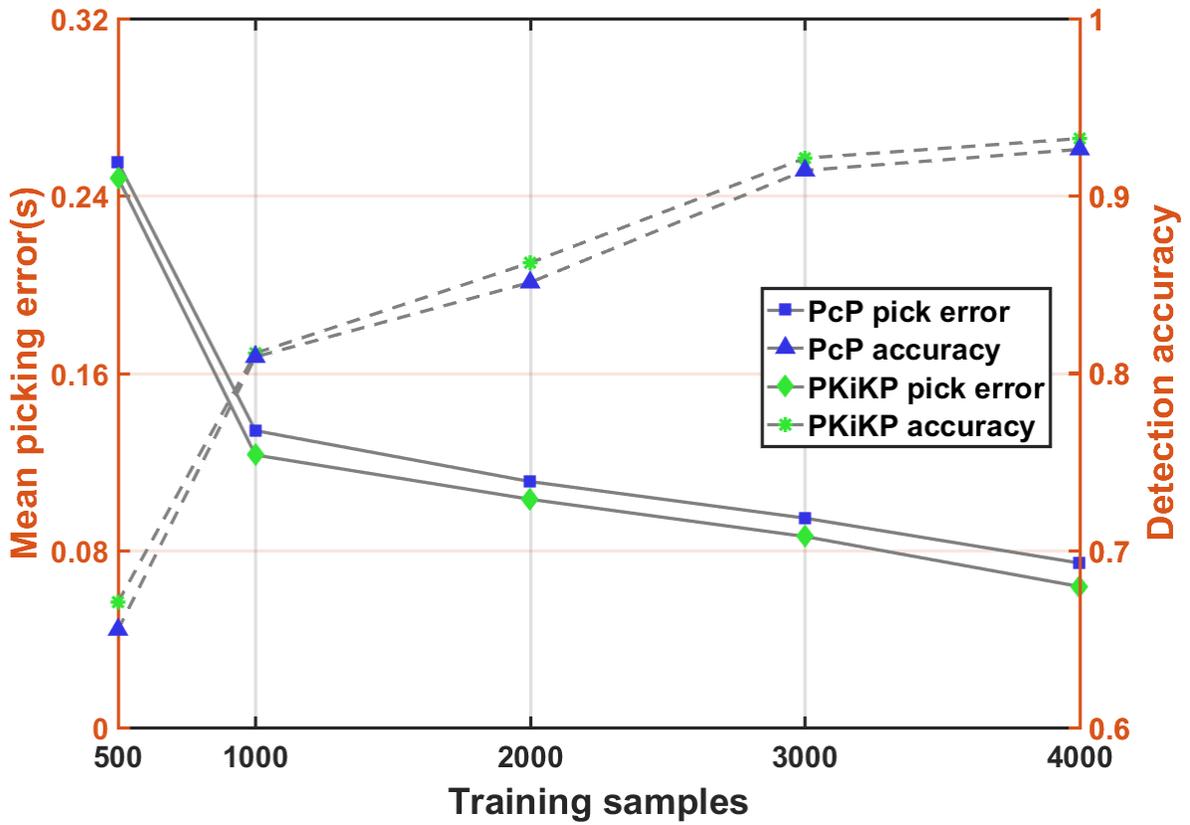

**Figure 10.** The effects of training samples on the detection accuracy and mean picking error of 100 testing samples.